\documentclass[conference,compsoc]{IEEEtran}
\ifCLASSOPTIONcompsoc
  \usepackage[nocompress]{cite}
\else
  \usepackage{cite}
\fi
\ifCLASSINFOpdf
\else
\fi
\usepackage{array}
\usepackage{multirow}
\usepackage{listings}
\usepackage{graphicx}
\usepackage{url}
\usepackage{xcolor}
\usepackage{listing}
\definecolor{my_white}{rgb}{1, 1, 1}                % Background
\definecolor{my_black}{rgb}{0, 0, 0}                % Code
\definecolor{my_blue}{rgb}{0, 0.125, 0.376}         % Keyword
\definecolor{my_green}{rgb}{0.12, 0.3, 0.17}        % Comment
\definecolor{my_violet}{rgb}{0.44, 0.16, 0.39}      % String

\lstdefinestyle{Bash}{
  backgroundcolor=\color{my_white},   % choose the background color
  basicstyle=\footnotesize,        % the size of the fonts that are used for the code
  breakatwhitespace=false,         % Automatic breaks should only happen at whitespace
  breaklines=true,                 % sets automatic line breaking
  captionpos=b,                    % sets the caption-position to bottom
  commentstyle=\color{my_green},   % comment style
  frame=single,	                   % adds a frame around the code
  keepspaces=true,                 % keeps spaces in text
  keywordstyle=\color{my_blue},    % keyword style
  language=bash,                   % the language of the code
  numbers=none,                    % Line numbers
  numbersep=5pt,                   % how far the line-numbers are from the code
  numberstyle=\tiny\color{my_black},   % the style that is used for the line-numbers
  rulecolor=\color{my_black},      % Rule color
  showspaces=false,                % underscores for spaces
  showstringspaces=false,          % underline spaces within strings only
  showtabs=false,                  % underscores for tabs
  stepnumber=1,                    % the step between two line-numbers
  stringstyle=\color{my_violet},   % string literal style
  tabsize=2,	                   % sets default tabsize to 2 spaces
  morekeywords={git, sudo, apt, get, pip2, okular, wxparaver},             % Extra keywords
  title=\lstname                   % show the filename of files included
}

\lstdefinestyle{PyCOMPSs}{
  backgroundcolor=\color{my_white},   % choose the background color
  basicstyle=\footnotesize,        % the size of the fonts that are used for the code
  breakatwhitespace=false,         % Automatic breaks should only happen at whitespace
  breaklines=true,                 % sets automatic line breaking
  captionpos=None,                    % sets the caption-position to bottom
  commentstyle=\color{my_green},   % comment style
  frame=single,	                   % adds a frame around the code
  keepspaces=true,                 % keeps spaces in text
  keywordstyle=\color{my_blue},    % keyword style
  language=Python,                 % the language of the code
  numbers=none,                    % Line numbers
  numbersep=5pt,                   % how far the line-numbers are from the code
  numberstyle=\tiny\color{my_black},   % the style that is used for the line-numbers
  rulecolor=\color{my_black},      % Rule color
  showspaces=false,                % underscores for spaces
  showstringspaces=false,          % underline spaces within strings only
  showtabs=false,                  % underscores for tabs
  stepnumber=1,                    % the step between two line-numbers
  stringstyle=\color{my_violet},   % string literal style
  tabsize=2,	                   % sets default tabsize to 2 spaces
  morekeywords={@task, @constraint, @parallel, compss_wait_on},             % Extra keywords
}

\begin{document}
\title{AutoParallel: A Python module for automatic parallelization \\and distributed execution of affine loop nests}

\author{
\IEEEauthorblockN{Cristian Ramon-Cortes}
\IEEEauthorblockA{Barcelona Supercomputing Center\\
Barcelona, Spain\\
Email: cristian.ramoncortes@bsc.es}

\and

\IEEEauthorblockN{Ramon Amela}
\IEEEauthorblockA{Barcelona Supercomputing Center\\
Barcelona, Spain\\
Email: ramon.amela@bsc.es}

\and

\IEEEauthorblockN{Jorge Ejarque}
\IEEEauthorblockA{Barcelona Supercomputing Center\\
Barcelona, Spain\\
Email: jorge.ejarque@bsc.es}

\and

\IEEEauthorblockN{Philippe Clauss}
\IEEEauthorblockA{INRIA\\
ICube Lab. - Universit\'{e} de Strasbourg \\
Strasbourg, France\\
Email: philippe.clauss@inria.fr}

\and

\IEEEauthorblockN{Rosa M. Badia}
\IEEEauthorblockA{Barcelona Supercomputing Center\\
Consejo Superior de Investigaciones Cientificas (CSIC) \\
Barcelona, Spain\\
Email: rosa.m.badia@bsc.es}
}


% compsoc conferences compared to traditional conferences), use this
%Telephone: (800) 555--1212, Fax: (888) 555--1212}


\maketitle

\begin{abstract}

The last improvements in programming languages, programming models, and frameworks have focused on abstracting the users from many programming issues. Among others, recent programming frameworks include simpler syntax, automatic memory management and garbage collection, which simplifies code re-usage through library packages, and easily configurable tools for deployment. For instance, Python has risen to the top of the list of the programming languages~\cite{top_langs} due to the simplicity of its syntax, while still achieving a good performance even being an interpreted language. Moreover, the community has helped to develop a large number of libraries and modules, tuning them to obtain great performance.

However, there is still room for improvement when preventing users from dealing directly with distributed and parallel computing issues. This paper proposes and evaluates AutoParallel, a Python module to automatically find an appropriate task-based parallelization of affine loop nests to execute them in parallel in a distributed computing infrastructure. This parallelization can also include the building of data blocks to increase task granularity in order to achieve a good execution performance. Moreover, AutoParallel is based on sequential programming and only contains a small annotation in the form of a Python decorator so that anyone with little programming skills can scale up an application to hundreds of cores. 

%It relies on PLUTO to parallelize affine loop nests and taskifies the obtained code so that PyCOMPSs can distributedly execute it in any underlying infrastructure (clusters, clouds, and containers). 
%Furthermore, we extend AutoParallel to automatically build data blocks so that the tasks' granularity is more suitable for PyCOMPSs. Hence, improving the execution performance.

\end{abstract} 

%\keywords{HPC, Python, Big Data, Linear Algebra}

% \fi
\IEEEpeerreviewmaketitle

% TEXT
\section{Introduction}

%----------------------------------------------------------------------------------------
% GENERAL INTRODUCTION

Computer simulations have become more and more crucial to both theoretical and experimental studies in many different fields, such as structural mechanics, chemistry, biology, genetics, and even sociology. Several years ago, small simulations (with up to several cores or even several nodes within the same grid) were enough to fulfill the scientific community requirements and thus, the experts of each field were capable of programming and running them. However, nowadays, simulations requiring hundreds or thousands of cores are widely used and, to this point, efficiently programming them becomes a challenge even for computer scientists. 
On the one hand, interdisciplinary teams have become popular, with field experts and computer scientists joining their forces together to keep their research at the forefront. On the other hand, programming languages have made a considerable effort to ease the programmability while maintaining acceptable performance. In this sense, Python~\cite{van2003python} has risen to the top language for nonexperts~\cite{top_langs}, being easy to program while maintaining good performance trade-off and having a large number of third-party libraries available. Similarly, Go~\cite{web:go_lang} has also gained some momentum thanks to its portability, reliability, and ease of concurrent programming, although it is still in its early stages.

%----------------------------------------------------------------------------------------
% PAPER CONTRIBUTION

Even if some great efforts have been accomplished for programming frameworks to ease the development of distributed applications, we go one step further with AutoParallel: a Python module to automatically parallelize applications and execute them in distributed environments. Our philosophy is to ease the development of parallel and distributed applications so that anyone with little programming skills can scale up an application to hundreds of cores. In this sense, AutoParallel is based on sequential programming and only requires a single Python decorator that frees the user from manually taskifying the original code. It relies on PLUTO~\cite{pluto1} to parallelize affine loop nests and taskifies the obtained code so that PyCOMPSs can distributedly execute it using any underlying infrastructure (clusters, clouds, and containers). Moreover, to avoid single instruction tasks, AutoParallel also includes an optional feature to increase the tasks' granularity by automatically building data blocks from PLUTO tiles.

%----------------------------------------------------------------------------------------
% PAPER STRUCTURE

The rest of the paper is organized as follows. Section~\ref{state_of_the_art} describes the state of the art. Section~\ref{tech_background} presents PyCOMPSs and PLUTO. Next, Section~\ref{architecture} describes the architecture of AutoParallel and Section~\ref{experimentation} presents its performance results. Finally, Section~\ref{conclusions} concludes the paper and gives some guidelines for future work.

\section{State of the Art}
\label{state_of_the_art}

Nowadays simulations are run in distributed environments and, although Python has become a reference programming language, there is still much work to do to ease parallel and distributed computing issues. In this concern, Python can provide parallelism at three levels. First, parallelism can be achieved internally through many libraries such as NumPy~\cite{walt2011numpy} and SciPy~\cite{jones2014scipy}, which offer vectorized data structures and numerical routines that automatically map operations on vectors and matrices to the BLAS~\cite{web:blas} and LAPACK~\cite{anderson1999lapack} functions; executing the multi-threaded BLAS version (using OpenMP~\cite{dagum1998openmp} or TBB~\cite{web:tbb}) when present in the system. Notice that, although parallelism is completely transparent for the application user, parallel libraries only benefit from intra-node parallelism and cannot run across different nodes.

Secondly, many modules can explicitly provide parallelism. The multiprocessing module provides~\cite{parallelprocessing} support for the spawning of processes in SMP machines using an API similar to the threading module, with explicit calls for creating processes. In addition, the Parallel Python (PP) module~\cite{parallelpython} provides mechanisms for parallel execution of Python codes, with an API that includes specific functions for specifying the number of workers to be used, submitting the jobs for execution, getting the results from the workers, etc. Also, the mpi4py~\cite{mpi4py} library provides a binding of MPI for Python which allows the programmer to handle parallelism both inter-node and intra-node. However, in all cases, the burden of parallelism specific issues is assigned to the programmer.

Finally, other libraries and frameworks enable Python distributed and multi-threaded computations such as Dask~\cite{dask}, PySpark~\cite{web:pyspark}, and PyCOMPSs~\cite{pycompss, pycompss_ogst}. Dask is a native Python library that allows both the creation of custom DAG's and the distributed execution of a set of operations on NumPy and pandas~\cite{mckinneypandas} objects. PySpark is a binding to the widely extended framework Spark~\cite{zaharia2010spark}. PyCOMPSs is a task-based programming model that offers an interface on Python that follows the sequential paradigm. It enables the parallel execution of tasks by means of building, at execution time, a data dependency graph for the tasks that compose an application. The syntax of PyCOMPSs is minimal, using decorators to enable the programmer to identify methods as tasks and a small API for synchronization. PyCOMPSs relies on a runtime that can exploit the inherent parallelism at task level and execute the application using a distributed parallel platform (clusters, clouds, and containers).

\section{Technical Background}
\label{tech_background}

%----------------------------------------------------------------------------------------
% GENERAL COMMENTS BEFORE SUBSECTIONS

This section provides a general overview of the satellite frameworks that directly interact with the AutoParallel module: PyCOMPSs and PLUTO. It also highlights some of their features that are crucial for their integration.

%----------------------------------------------------------------------------------------
\subsection{PyCOMPSs}
\label{pycompss}

COMPSs~\cite{compss_softwareX, compss_servicess} is a task-based programming model that aims to ease the development of parallel applications, targeting distributed computing platforms. It relies on its runtime to exploit the inherent parallelism of the application at execution time by detecting the task calls and the data dependencies between them.

The COMPSs runtime natively supports Java applications but also provides bindings for Python and C/C++. Precisely, the Python binding is known as PyCOMPSs. All the bindings are supported through a \textit{binding-commons} layer which focuses on enabling the functionalities of the runtime to other languages. It is written in C and has been designed as an API with a set of defined functions to communicate with the runtime through the JNI~\cite{liang1999jni}.

\begin{figure}[!htb]
  \centering
  \includegraphics[width=0.7\linewidth]{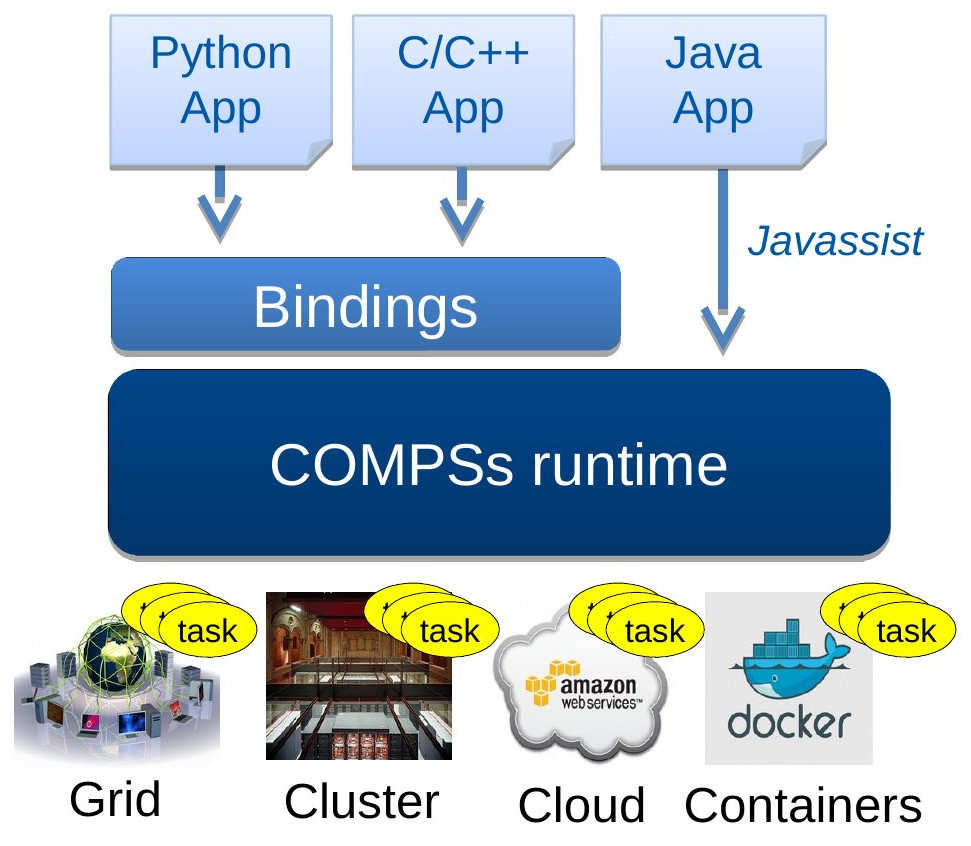}
  \caption{COMPSs overview}
  \label{fig:compss_overview}
\end{figure}

As shown in Figure~\ref{fig:compss_overview}, the COMPSs runtime allows applications to be executed on top of different infrastructures (such as multi-core machines, grids, clouds or containers~\cite{compss_orchestration}) without modifying a single line of the application code. Thanks to the different connectors, the runtime is capable of handling all the underlying infrastructure so that the user only defines the tasks. It also provides fault-tolerant mechanisms for partial failures (with job resubmission and reschedule when tasks or resources fail), has a live monitoring tool through a built-in web interface, supports instrumentation using the Extrae~\cite{web:extrae} tool to generate post-mortem traces that can be analyzed with Paraver~\cite{web:paraver}, has an Eclipse IDE, and has pluggable cloud connectors and task schedulers. 

Additionally, the programming model is based on sequential programming which means that users do not need to deal with any parallelization and distribution issue such as thread creation, synchronization, data distribution, messaging or fault-tolerance. Instead, application developers only select which methods must be considered as tasks, and the runtime spawns them asynchronously on a set of resources instead of executing them locally and sequentially.

\subsubsection{PyCOMPSs Programming Model} ~\newline

Regarding programmability, tasks are identified by inserting annotations in the form of Python decorators. These annotations are inserted at method level and indicate that invocations to a given method should become tasks at execution time. The \texttt{@task} decorator also contains information about the directionality of the method parameters specifying if a given parameter is read (IN), written (OUT) or both read and written in the method (INOUT).

Figure~\ref{lst:sample_task} shows an example of a task annotation. The parameter \texttt{c} has direction INOUT, and parameters \texttt{a} and \texttt{b} are set to the default direction IN. The directionality tags are used at execution time to derive the data dependencies between tasks and are applied at an object level, taking into account its references to identify when two tasks access the same object.

Additionally to the \texttt{@task} decorator, the \texttt{@constraint} decorator can be optionally defined to indicate some task hardware or software requirements. Continuing with the previous example, the task constraint \texttt{ComputingUnits} tells the runtime how many CPUs are consumed by each task execution. The available resources are defined by the system administrator in a separated XML configuration file. Other constraints that can be defined refer to the processor architecture, memory size, disk storage, operating system or available libraries.

%\begin{figure}[!htb]
%  \centering
%  \small
%  \begin{minted}{python}
%    @constraint(ComputingUnits="$CUS")
%    @task(c=INOUT)
%    def multiply(a, b, c):
%        c += a * b
%  \end{minted}
%  \caption{Sample task annotation}
%  \label{lst:sample_task}
%\end{figure}

\begin{figure}[!htb]
  \centering
\begin{lstlisting}[style=PyCOMPSs, basicstyle=\scriptsize]
@constraint(ComputingUnits="$CUS")
@task(c=INOUT)
def multiply(a, b, c):
    c += a * b
\end{lstlisting}
  \caption{Sample task annotation}
  \label{lst:sample_task}
\end{figure}

A tiny synchronization API completes the PyCOMPSs syntax. As shown in Figure~\ref{lst:sample_call}, the API function \texttt{compss\_wait\_on} waits for the completion of all the tasks modifying the \texttt{result}'s value and brings the final value to the node executing the main program. Then, the execution of the main program is resumed. Given that PyCOMPSs is used mostly in distributed environments, synchronization implies a data transfer from remote storage or memory space to the node executing the main program.

%\begin{figure}[!htb]
%  \centering
%  \small
%  \begin{minted}{python}
%for block in data:
%    partial_res = wordcount_task(block)
%    reduce_task(result, partial_res)
%final_result = compss_wait_on(result)
%  \end{minted}
%  \caption{Sample call to synchronization API}
%  \label{lst:sample_call}
%\end{figure}

\begin{figure}[!htb]
  \centering
\begin{lstlisting}[style=PyCOMPSs, basicstyle=\scriptsize]
for block in data:
    partial_res = wordcount_task(block)
    reduce_task(result, partial_res)
final_result = compss_wait_on(result)
\end{lstlisting}
  \caption{Sample call to synchronization API}
  \label{lst:sample_call}
\end{figure}

\subsection{PLUTO}
\label{pluto}

\begin{figure}[!htb]
  \centering
  \includegraphics[width=\linewidth]{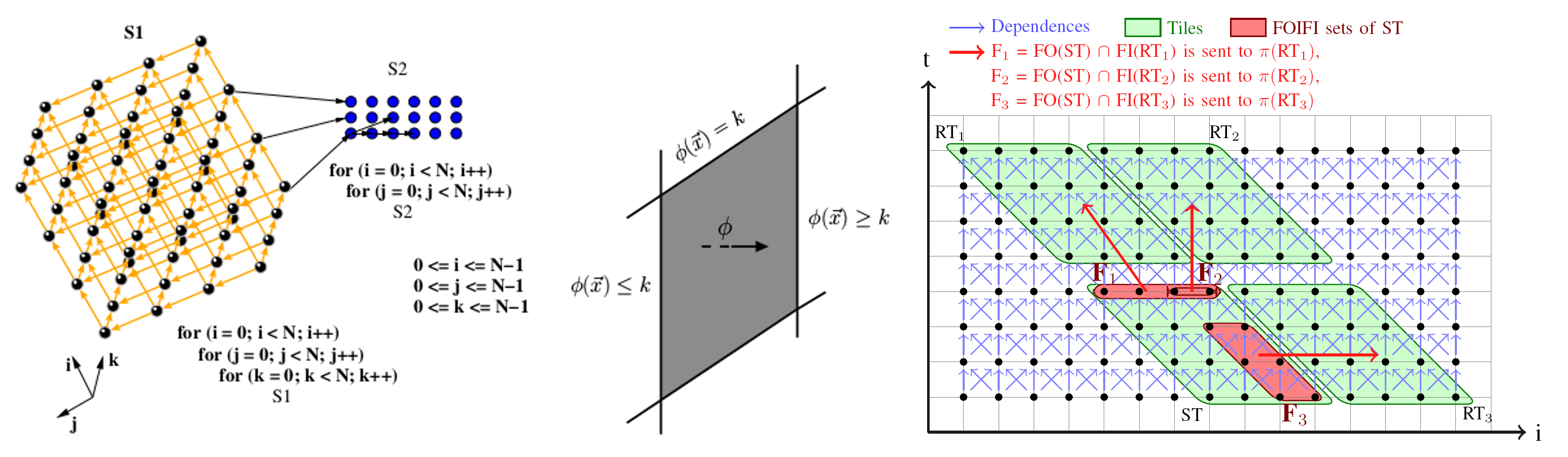} \\
  \caption[Pluto overview]{Pluto overview. Source: Pluto's official website~\cite{web:pluto}}
  \label{fig:pluto_overview}
\end{figure}

Many compute-intensive scientific applications spend most of their execution time running nested loops. The Polyhedral Model~\cite{polyhedral_model} provides a powerful mathematical abstraction to analyze and transform loop nests in which the data access functions and loop bounds are affine combinations (linear combinations with a constant) of the enclosing loop iterators and parameters. As shown in Figure~\ref{fig:pluto_overview}, this model represents the instances of the loop nests' statements as integer points inside a polyhedron, where inter and intra-statement dependencies are characterized as a dependency polyhedron. Combining this representation with Linear Algebra and Integer Linear Programming, it is possible to reason about the correctness of a sequence of complex optimizing and parallelizing loop transformations.

PLUTO~\cite{web:pluto, pluto1} is an automatic parallelization tool based on the Polyhedral model to optimize arbitrarily nested loop sequences with affine dependencies. At compile time, it analyses C source code to optimize and parallelize affine loop-nests and automatically generate OpenMP C parallel code for multi-cores. Although the tool is fully automatic, many options are available to tune tile sizes, unroll factors, and outer loop fusion structure.

\begin{figure}[!htb]
  \centering
  \includegraphics[width=\linewidth]{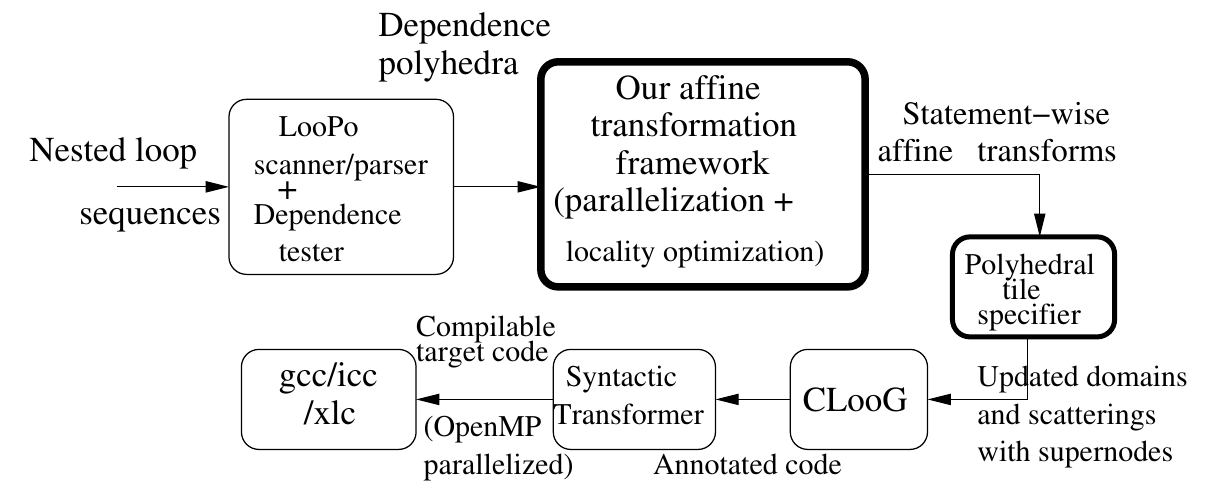} \\
  \caption[Pluto source-to-source transformation]{Pluto source-to-source transformation. Source: ~\cite{pluto2}}
  \label{fig:pluto_internals}
\end{figure}

As shown in Figure~\ref{fig:pluto_internals}, PLUTO internally translates the source code to an intermediate OpenScop~\cite{openscop} representation using CLAN~\cite{clan}. Next, it relies on the Polyhedral Model to find affine transformations for coarse-grained parallelism, data locality, and efficient tiling. Finally, PLUTO generates the OpenMP C code from the OpenScop representation using CLooG~\cite{cloog}. We must highlight that the generated code is also optimized for data locality and made amenable to auto-vectorization.

\subsubsection{Loop Tiling} ~\newline

Among many other options, PLUTO can tile code by specifying the \verb|--tile| option. In general terms, as shown in Figure~\ref{lst:tiling_example}, tiling a loop of given size \verb|N| results in a division of the loop in \verb|N/T| repeatable parts of size \verb|T|. For instance, this is suitable when fitting loops into the L1 or L2 caches or, in the context of this paper, when building the data blocks to increase the tasks' granularity. 

%\begin{figure}[!htb]
%  \small
%  \begin{minted}{python}
%# Original loop       # Tiled loop
%for i in range(N):    for i in range(N/T):
%    print(i)              for t in range(T):
%                              print(i*T + t)
%  \end{minted}
%  \caption{Tiling example}
%  \label{lst:tiling_example}
%\end{figure}

\begin{figure}[!htb]
  \centering
\begin{lstlisting}[style=PyCOMPSs, basicstyle=\scriptsize]
# Original loop       # Tiled loop
for i in range(N):    for i in range(N/T):
    print(i)              for t in range(T):
                              print(i*T + t)

\end{lstlisting}
  \caption{Tiling example}
  \label{lst:tiling_example}
\end{figure}

Along with this option, users can let PLUTO set the tile sizes automatically using a rough heuristic, or manually define them in a \verb|tile.sizes| file. This file must contain one tile size on each line and as many tile sizes as the loop nest depth.

In the context of parallel applications, tile sizes must be fine-tuned for each application so that they maximize locality while making sure there are enough tiles to keep all cores busy.

\section{Architecture}
\label{architecture}

%----------------------------------------------------------------------------------------

The framework proposed in this paper eases the development of distributed applications by letting users program their application in a standard sequential fashion. It is developed on top of PyCOMPSs and PLUTO. When automatically parallelizing sequential applications, users must only insert an annotation on top of the potentially parallel functions to activate the AutoParallel module. Next, the application can be launched using PyCOMPSs.

\begin{table}[!htb]
  \centering
  \renewcommand{\arraystretch}{1.2}
  \begin{tabular}{| m{3cm} | m{0.8cm} | m{3cm} | }
    \hline
    \textbf{Flag} & \textbf{Default Value} & \textbf{Description} \\ \hline \hline
    \verb|pluto_extra_flags| & None & List of flags for the internal PLUTO command \\ \hline
    \verb|taskify_loop_level| & 0 & Taskification loop depth (see Section~\ref{loop_taskification}) \\ \hline
    \verb|force_autogen| & True & When set to False, loads a previously generated code \\ \hline
    \verb|generate_only| & False & When set to True, only generates the parallel version of the code \\
    \hline
  \end{tabular}
  \caption{List of flags for the \texttt{@parallel()} decorator}
  \label{tab:flags}
\end{table}

Following the same approach than PyCOMPSs, we have included a new decorator \texttt{@parallel()} to specify which methods should be automatically parallelized at runtime. Notice that functions using this decorator must contain affine loops so that the module can propose a parallelization. Otherwise, the source code will remain intact. Table~\ref{tab:flags} shows the valid flags for the decorator.

\begin{figure}[!htb]
  \centering
  \includegraphics[width=0.8\linewidth]{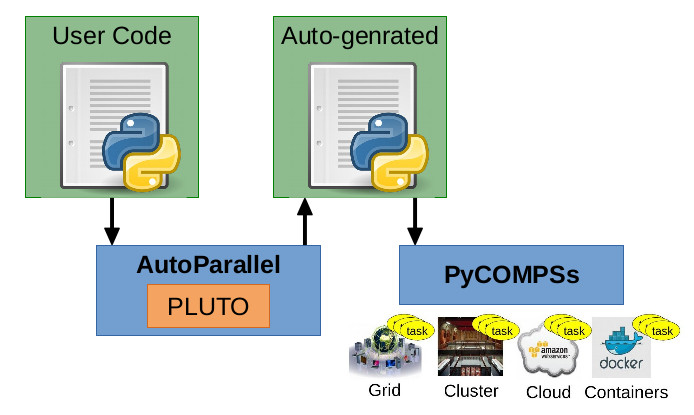}
  \caption{AutoParallel Module Overview}
  \label{fig:autoparallel_overview}
\end{figure}

As shown in Figure~\ref{fig:autoparallel_overview}, the AutoParallel Module analyzes the user code searching for \texttt{@parallel()} annotations. Essentially, when found, the module calls PLUTO to generate its parallelization and substitutes the user code by a newly generated code. Once all annotations have been processed, the new tasks are registered into PyCOMPSs, and the execution continues as a regular PyCOMPSs application (as described in Section~\ref{pycompss}). Finally, when the application has ended, the generated code is stored in a (\texttt{\_autogen.py}) file and the user code is restored. 

%----------------------------------------------------------------------------------------

\subsection{AutoParallel Module}

\begin{figure}[!htb]
  \centering
  \includegraphics[width=0.75\linewidth]{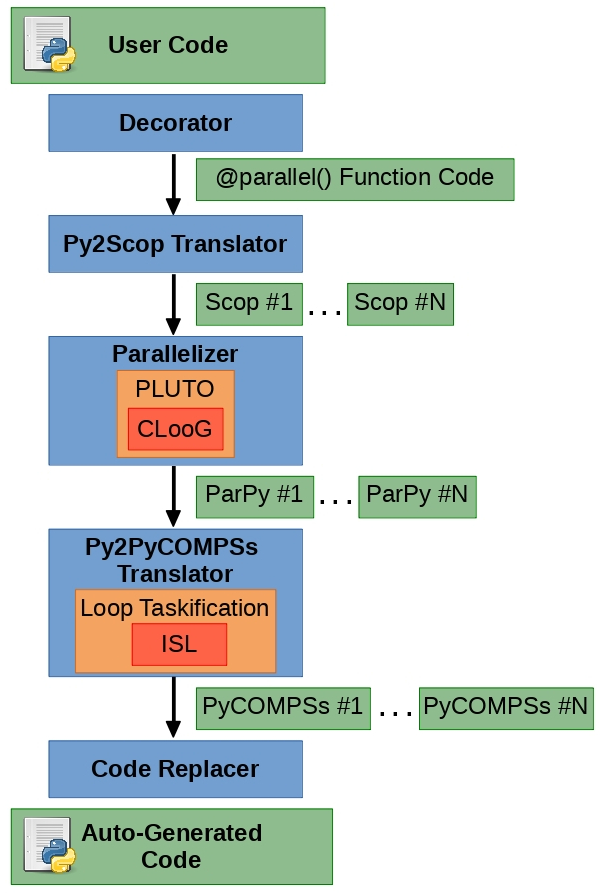}
  \caption{AutoParallel Module Internals}
  \label{fig:autoparallel_internals}
\end{figure}

Next, we describe the five components of the AutoParallel module. For the sake of clarity, Figure~\ref{fig:autoparallel_internals} shows the relationship between all the components and its expected inputs and outputs. 

\begin{itemize}
  \item \textbf{Decorator} Implements the \texttt{@parallel()} decorator to detect functions that the user has marked as potentially parallel 
  \item \textbf{Python To OpenScop Translator} For each affine loop nest detected in the user function, builds a Python Scop object representing it that can be bulked into an OpenScop format file.
  \item \textbf{Parallelizer} Returns the Python code resulting from parallelizing an OpenScop file. Since Python does not have any standard regarding parallel annotations, the parallel loops are annotated using comments with OpenMP syntax.
  \item \textbf{Python to PyCOMPSs Translator} Converts an annotated Python code into a PyCOMPSs application by inserting the necessary task annotations and data synchronizations. This component also performs loop taskification if it is enabled (see Section~\ref{loop_taskification} for more details). 
  \item \textbf{Code Replacer} Replaces each loop nest in the initial user code by the auto-generated code so that PyCOMPSs can execute the code in a distributed computing platform. When the application has finished, it restores the user code and saves the auto-generated code in a separated file.
\end{itemize}

For instance, Figure~\ref{fig:example_lt_src} shows the relevant parts of an Embarrassingly Parallel application with the main function annotated with the \texttt{@parallel()} decorator that contains two nested loops. 

\begin{figure}[!htb]
  \centering
\begin{lstlisting}[style=PyCOMPSs, basicstyle=\scriptsize]
# Main Function
from pycompss.api.parallel import parallel
@parallel()
def ep(mat, n_size, m_size, c1, c2):
    for i in range(n_size):
        for j in range(m_size):
            mat[i][j] = compute(mat[i][j], c1, c2)
            
# MAIN
if __name__ == "__main__":
    # Parse arguments
    ...
    # Initialize matrices
    MAT = initialize_matrix(NSIZE, MSIZE)
    # Begin computation
    ep(MAT, NSIZE, MSIZE, COEF1, COEF2)
\end{lstlisting}
  \caption{EP example: user code}
  \label{fig:example_lt_src}
\end{figure}

\begin{figure}[!htb]
  \centering
\begin{lstlisting}[style=PyCOMPSs, basicstyle=\scriptsize]
# [COMPSs Autoparallel] Begin Autogenerated code
@task(var2=IN, c1=IN, c2=IN, returns=1)
def S1(var2, c1, c2):
    return compute(var2, c1, c2)

def ep(mat, n_size, m_size, c1, c2):
    if m_size >= 1 and n_size >= 1:
        lbp = 0
        ubp = m_size - 1
        for t1 in range(lbp, ubp + 1):
            lbv = 0
            ubv = n_size - 1
            for t2 in range(lbv, ubv + 1):
                mat[t2][t1]=S1(mat[t2][t1],c1,c2)
    compss_barrier()
# [COMPSs Autoparallel] End Autogenerated code
\end{lstlisting}
  \caption{EP example: auto-generated code without loop taskification}
  \label{fig:example_lt_single}
\end{figure}

In addition, Figure~\ref{fig:example_lt_single} shows its parallelization. Notice that the source code contains two nested loops with a single inner statement. It contains a task definition (with its data dependencies) that matches the original loop statement, a new main loop nest proposed by PLUTO (which exploits the inherent parallelism available in the original code), and a final barrier used as synchronization point.

\subsection{Loop Taskification}
\label{loop_taskification}

Many compute-intensive scientific applications are not designed as block computations, and thus, the tasks proposed by the AutoParallel module are single statements. Although this can be harmless in tiny parallel environments, it leads to poor performances when executed using large distributed environments since the tasks' granularity is not large enough to surpass the overhead of transferring the task definition, and the input and output data. To face this issue, we have extended the \textit{Python to PyCOMPSs Translator} with support for loop taskification.

Essentially, loop taskification means processing the parallel code and converting into tasks all the loops of a certain depth of the loop nest. Since tasks may use N-dimensional arrays, this also implies to create the necessary data blocks (chunks) for each callee and revert them after the task execution. Continuing with the previous example, Figure~\ref{fig:example_lt_multi} shows its parallelization with loop taskification.

\begin{figure}[!htb]
  \centering
\begin{lstlisting}[style=PyCOMPSs, basicstyle=\scriptsize]
# [COMPSs Autoparallel] Begin Autogenerated code
@task(lbv=IN, ubv=IN, c1=IN, c2=IN, returns="LT2_args_size")
def LT2(lbv, ubv, c1, c2, *args):
    global LT2_args_size
    var1, = ArgUtils.rebuild_args(args)
    for t2 in range(0, ubv + 1 - lbv):
        var1[t2] = S1_no_task(var1[t2], c1, c2)
    return ArgUtils.flatten_args(var1)

def S1_no_task(var2, c1, c2):
    return compute(var2, c1, c2)

def ep(mat, n_size, m_size, c1, c2):
    if m_size >= 1 and n_size >= 1:
        lbp = 0
        ubp = m_size - 1
        for t1 in range(lbp, ubp + 1):
            lbv = 0
            ubv = n_size - 1
            # Chunk creation and flattening
            LT2_aux_0 = [mat[t2][t1] for ...]
            LT2_au = ArgUtils()
            global LT2_args_size
            LT2_flat, LT2_args_size 
                = LT2_au.flatten(LT2_aux_0)
            # Task call
            LT2_ret = LT2(lbv, ubv, c1, c2,
                                *LT2_flat)
            # Rebuild and re-assign
            LT2_aux_0, = LT2_au.rebuild(LT2_ret)
            ...
    compss_barrier()
# [COMPSs Autoparallel] End Autogenerated code
\end{lstlisting}
  \caption{EP example: auto-generated code with loop taskification}
  \label{fig:example_lt_multi}
\end{figure}

Notice that the generated code with loop taskification is significantly more complex. The code defines a task containing the inner-most loop of the original code' statements (only one in this case) and all its data dependencies accordingly. The main loop is also modified by chunking the necessary data for the task execution, flattening the data as a single dimension list, calling the task, rebuilding the chunks after the task callee, and re-assigning the values to the original variables. Notice that this last assignment is performed by pre-calculating the number of expected parameters and using PyCOMPSs Future Objects to avoid any synchronization in the master code (objects are only synchronized and transferred when required by a task execution). The end of the main code also includes a barrier as a synchronization point. 

% \begin{figure}[!htb]
%   \centering
%   \lstinputlisting[style=PyCOMPSs]{listings/example_lt_multi_tile.py}
%   \caption{EP example: auto-generated code with PLUTO tiling and loop taskification}
%   \label{fig:example_lt_multi_tile}
% \end{figure}

However, taskification may annihilate all the potential parallelism of the application (for instance, by setting \verb|taskify_loop_level=2| in the previous example). To prevent this, we have integrated PLUTO's tiling transformation so that the loop taskification can be achieved on tiles. %Figure~\ref{fig:example_lt_multi_tile} shows the previous Embracingly Parallel example combining PLUTO's tiling and the loop taskification. 

%----------------------------------------------------------------------------------------
\subsection{Python Extension for CLooG}

As described in Section~\ref{pluto}, PLUTO operates internally with the OpenScop format. It relies on CLAN to translate input code from C/C++ or Fortran to OpenScop, and on CLooG to translate output code from OpenScop to C/C++ or Fortran.

Since we are targeting Python code, a translation from Python to OpenScop is required before calling PLUTO, and another translation from OpenScop to Python is required at the end. For the input, we have chosen to manually translate the code using the \textit{Python To OpenScop Translator} component since CLAN is not adapted for supporting other language and PLUTO may take OpenScop format as input. We have extended CLooG so that the written output code is directly in Python: the language options were extended, and the Pretty Printer was modified in order to translate every OpenScop statement in its equivalent Python format. Since Python does not have any standard regarding parallel annotations, the parallel loops are annotated with comments in OpenMP syntax.

\section{Experimentation}
\label{experimentation}

%----------------------------------------------------------------------------------------
% GENERAL INTRODUCTION

\subsection{Computing Infrastructure}

Results presented in this section have been obtained using the MareNostrum IV Supercomputer located at the Barcelona Supercomputing Center (BSC).

We have used PyCOMPSs version 2.3.rc1807 (available at~\cite{github:compss}), PLUTO version 0.11.4, CLooG version 0.19.0, and AutoParallel version 0.2 (available at~\cite{github:autoparallel}). We have also used Intel\textregistered Python 2.7.13, Intel\textregistered MKL 2017, Java OpenJDK 8 131, GCC 7.2.0, and Boost 1.64.0.

All the benchmark codes used for this experimentation are also available at~\cite{github:experimentation}.

\subsubsection*{MareNostrum IV}

The MareNostrum IV begun operating at the end of June 2017. Its current peak performance is 11.15 Petaflops, ten times more than its previous version, MareNostrum III. The supercomputer is composed by 3456 nodes, each of them with two Intel\textregistered \ Xeon Platinum 8160 (24 cores at 2,1 GHz each). It has 384.75 TB of main memory, 100Gb Intel\textregistered Omni-Path Full-Fat Tree Interconnection, and 14 PB of disk storage~\cite{web:mn4}.

%----------------------------------------------------------------------------------------
% EXPERIMENTS

%--------------------------------
% Blocked Apps
\subsection{Blocked Applications}

The first set of experiments has been designed to compare the application's code automatically generated by the AutoParallel module (\textit{autoparallel} version) against the one written by a PyCOMPSs expert user (\textit{userparallel} version). To this end, we have used the LU, Cholesky, and QR decompositions described and analyzed in our previous work~\cite{pycompss}.

In general terms, the matrices are chunked in smaller square matrices (known as \emph{blocks}) to distribute the data easily among the available resources so that the square blocks are the minimum entity to work with~\cite{gunnels2001flame}. Furthermore, the initialization is performed in a distributed way, defining tasks to initialize the matrix blocks. These tasks do not take into account the nature of the algorithm, and they are scheduled in a round robin manner. Next, all the computations are performed considering that the data is already located on a given node. 

Given a fixed matrix size, increasing the number of blocks increases the maximum parallelism of the application since blocks are the tasks' minimum work entities. On the other hand, increasing the block size increases the tasks' computational load which, at some point, will surpass the serialization and transfer overheads. Hence, the number of blocks and the block size for each application are a trade-off to fill all the available cores while maintaining acceptable performance.

Next subsections analyze each application in depth, with a figure showing the execution results that contains two plots. For both plots, the horizontal axis shows the number of worker nodes (with 48 cores each) used for each execution, the blue color is the \textit{userparallel} version and the green color is the \textit{autoparallel}. The top plot represents the mean, maximum, and minimum execution times over 10 runs and the bottom plot represents the speed-up of each version with respect to the \textit{userparallel} version running with a single worker.

% Cholesky
\subsubsection{Cholesky} ~\newline

The Cholesky factorization can be applied to Hermitian positive-defined matrices. This decomposition is a particular case of the LU factorization, obtaining two matrices of the form $U = L^t$. Our version of this application applies the right-looking algorithm~\cite{bientinesi2008cholesky} because it is more aggressive, meaning that in an early stage of the computation there are blocks of the solution that are already computed and all the potential parallelism is released as soon as possible. 

Table~\ref{tab:cholesky_code} analyses the \textit{userparallel} and \textit{autoparallel} versions in terms of code complexity, loop configuration, and number of different task types. The code complexity is measured using the Babelfish Tools~\cite{web:bblfsh-tools} and includes lines of code, cyclomatic complexity, and n-path. Although the generated code is not much larger than the original one, it is significantly more complex in terms of cyclomatic complexity and n-path. Notice that, although the \textit{autoparallel} version has three 3-depth loop nests instead of a single loop nest with four loops, the maximum loop depth remains the same (three). Furthermore, regarding the amount of task calls, \textit{userparallel} version includes three tasks (\verb|potrf|, \verb|solve_triangular|, and \verb|gemm|), while the \textit{autoparallel} version includes four tasks (that map to the previous operations plus an additional one to generate blocks initialized to zero).

\begin{table}[!htb]
  \centering
  \begin{tabular}{| m{1.2cm} | m{0.5cm} | m{0.5cm} | m{0.65cm} | m{0.5cm} | m{0.5cm} | m{0.6cm} | m{0.6cm} |}
    \hline
    \multirow{2}{*}{\textbf{Version}} & \multicolumn{3}{c|}{\textbf{Code Analysis}} & \multicolumn{3}{c|}{\textbf{Loops Analysis}} & \multirow{2}{0pt}{\textbf{Task Types}} \\ \cline{2-7}
    ~ &\textbf{LoC} &\textbf{CC} &\textbf{NPath} &\textbf{Main}	&\textbf{Total} &\textbf{Depth} &~ \\ \hline
    
    userparallel    & 220   & 26    & 112     & 1		& 4		& 3		& 3 \\ \hline
    autoparallel 	& 274   & 36    & 14576     & 3		& 9		& 3		& 4 \\ \hline
  \end{tabular}
  \caption{Cholesky code analysis}
  \label{tab:cholesky_code}
\end{table}

\begin{figure}[!htb]
  \centering
  \includegraphics[width=0.9\linewidth]{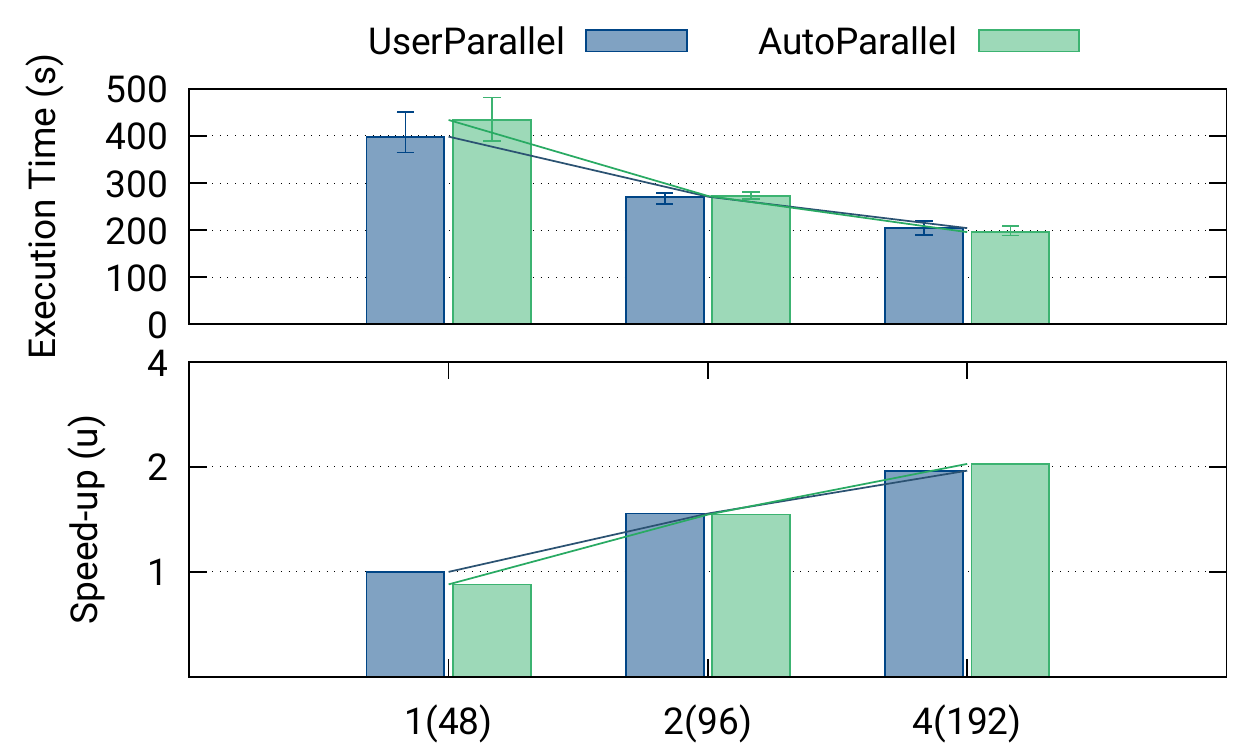}
  \caption{Cholesky decomposition: Execution times and speed-up}
  \label{fig:cholesky_scaling}
\end{figure}

Figure~\ref{fig:cholesky_scaling} shows the execution results of the Cholesky decomposition over a dense matrix of $65536 \times 65536$ elements decomposed in $32 \times 32$ blocks with $2048 \times 2048$ elements each. As explained at the beginning of this section, we have chosen 32 blocks because it is the minimum amount providing enough parallelism for 192 cores, and a bigger block size (e.g., $4096 \times 4096$) was impossible due to memory constraints. The speed-up of both versions is limited by the block-size due to the small task granularity, reaching 2 when using 4 workers. Although the \textit{userparallel} version spawns 6512 tasks and the \textit{autoparallel} version spawns 7008 tasks, the execution times and the overall performance of both versions are almost the same. This is due to the fact that the \textit{autoparallel} version spawns an extra task per iteration to initialize blocks to zero on the matrix's lower triangle that has no impact in the overall computation time.

% LU
\subsubsection{LU} ~\newline

For the LU decomposition, an approach without pivoting~\cite{golub1996matrix} has been the starting point. However, since this approach might be unstable in general~\cite{demmel1992blas}, some modifications have been included to increase the stability of the algorithm while keeping the block division and avoiding bringing an entire column into a single node.

As shown in Table~\ref{tab:lu_code}, both versions have two main loops with a maximum depth of three and a total of six nested loops. However, the \textit{autoparallel} version has 315\% more paths in the flow than the \textit{userparallel} because it has different optimization codes for different variable values. Regarding the task types, the \textit{userparallel} version contains calls to four different tasks: \verb|multiply|, \verb|invert_triangular|, \verb|dgemm|, and \verb|custom_lu|. On the other hand, the \textit{autoparallel} version generates twelve different task types because it generates one task type per statement in the original loop, even if the statement contains the same task call. For instance, the original LU contains four calls to the \verb|invert_triangular| function that are detected as different statements and converted to different task types.

\begin{table}[!htb]
  \centering
  \begin{tabular}{| m{1.2cm} | m{0.5cm} | m{0.5cm} | m{0.65cm} | m{0.5cm} | m{0.5cm} | m{0.6cm} | m{0.6cm} |}
    \hline
    \multirow{2}{*}{\textbf{Version}} & \multicolumn{3}{c|}{\textbf{Code Analysis}} & \multicolumn{3}{c|}{\textbf{Loops Analysis}} & \multirow{2}{0pt}{\textbf{Task Types}} \\ \cline{2-7}
    ~ &\textbf{LoC} &\textbf{CC} &\textbf{NPath} &\textbf{Main}	&\textbf{Total} &\textbf{Depth} &~ \\ \hline
    
    userparallel    & 238   & 35    & 79872     & 2		& 6		& 3		& 4 \\ \hline
    autoparallel 	& 320   & 39    & 331776     & 2		& 6		& 3		& 12 \\ \hline
  \end{tabular}
  \caption{LU code analysis}
  \label{tab:lu_code}
\end{table}

\begin{figure}[!htb]
  \centering
  \includegraphics[width=0.95\linewidth]{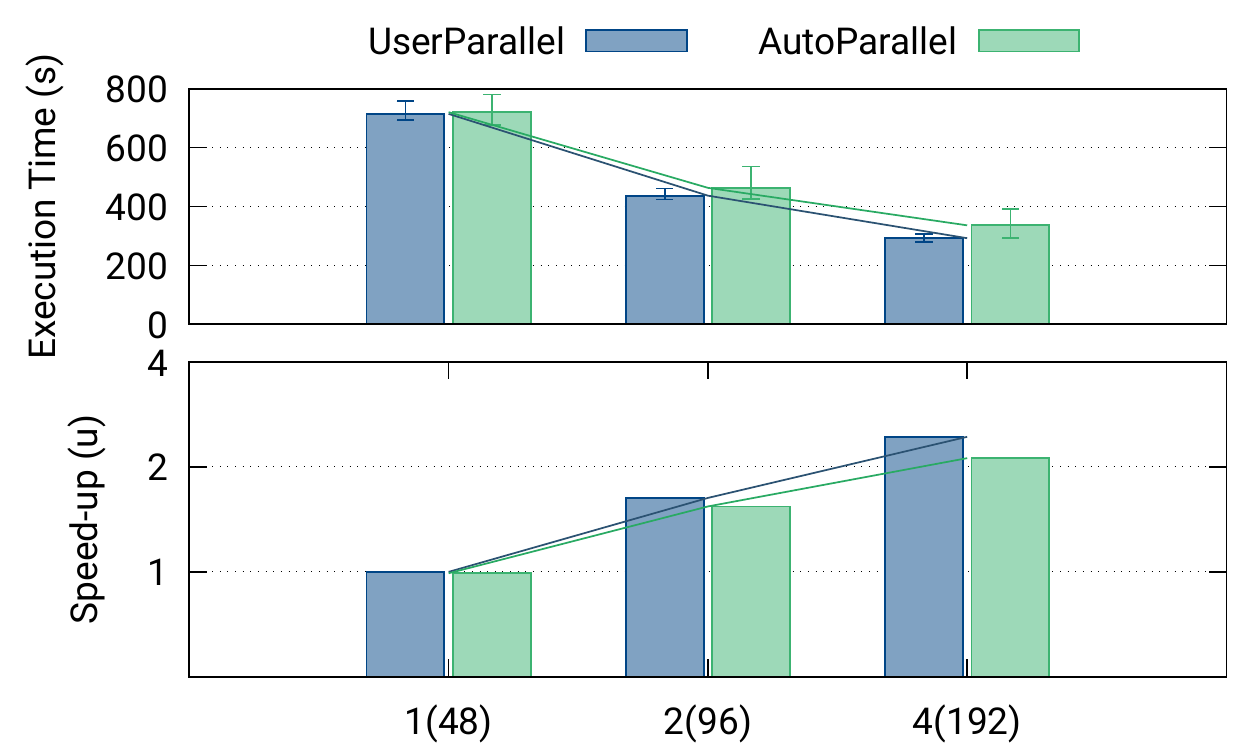}
  \caption{LU decomposition: Execution times and speed-up}
  \label{fig:lu_scaling}
\end{figure}

Figure~\ref{fig:lu_scaling} shows the execution results of the LU decomposition with a $49152 \times 49152$ dense matrix of $24 \times 24$ blocks with $2048 \times 2048$ elements each. As in the previous example, the overall performance is limited by the block size. This time the \textit{userparallel} version slightly outperforms the \textit{autoparallel} version; achieving, respectively, a 2.45 and 2.13 speed-up with 4 workers (192 cores).

\begin{figure*}[!htb]
  \centering
  \includegraphics[width=0.9\linewidth]{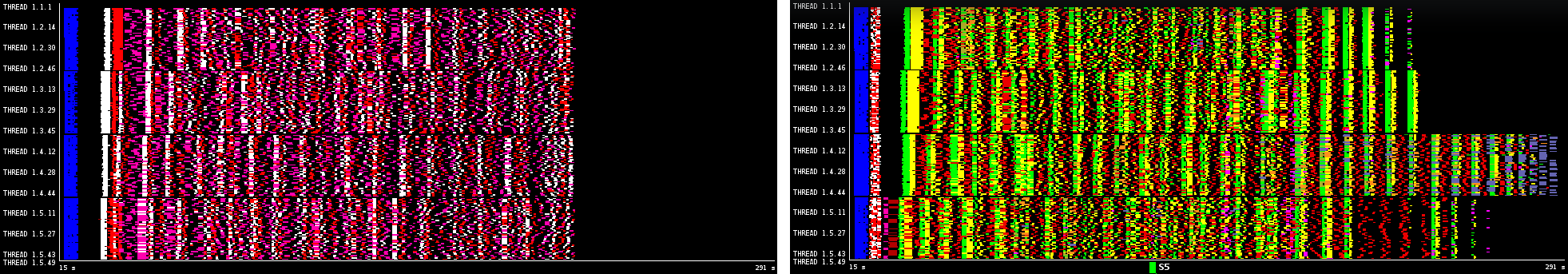}
  \caption{LU decomposition: Paraver trace. At left, \textit{userparallel} and, at right, the \textit{autoparallel} version}
  \label{fig:lu_trace}
\end{figure*}

Regarding the number of tasks, the \textit{userparallel} version spawns 14676 tasks while the \textit{autoparallel} version spawns 15227 tasks. This difference is due to the fact that the \textit{autoparallel} version initializes distributedly an intermediate zero matrix, while the \textit{userparallel} initializes it in the master memory. 

Figure~\ref{fig:lu_trace} shows a detailed Paraver trace of both versions running with 4 workers (192 cores). The \textit{autoparallel} version (right) is more coloured because it has more tasks although, as previously explained, they execute the same function in the end. Notice that the performance degradation of the \textit{autoparallel} version is due to the fact that the maximum parallelism is lost before the end of the execution. On the contrary, the \textit{userparallel} version maintains the maximum parallelism until the end of the execution.

% QR
\subsubsection{QR} ~\newline

Unlike traditional QR algorithms that use the Householder transformation, our implementation uses a method based on Givens rotations~\cite{quintana2008qr}. This way, data can be accessed by blocks instead of columns. 

The QR decomposition represents one of the most complex use cases in terms of data dependencies thus, having a high cyclomatic complexity. As shown in Table~\ref{tab:qr_code}, the generated code differs significantly in terms of loop configurations but not regarding code since the \textit{autoparallel} version has similar cyclomatic complexity, 34\% more lines of code, and 51\% more paths. Although the maximum loop depth remains the same, the auto-generated code has two main loops instead of one. Regarding the task types, while the \textit{userparallel} version has four tasks (namely \verb|qr|, \verb|dot|, \verb|little_qr|, and \verb|multiply_single_block|), the \textit{autoparallel} version has twenty different task types.

\begin{table}[!htb]
  \centering
  \begin{tabular}{| m{1.2cm} | m{0.5cm} | m{0.5cm} | m{0.6cm} | m{0.5cm} | m{0.5cm} | m{0.6cm} | m{0.6cm} |}
    \hline
    \multirow{2}{*}{\textbf{Version}} & \multicolumn{3}{c|}{\textbf{Code Analysis}} & \multicolumn{3}{c|}{\textbf{Loops Analysis}} & \multirow{2}{0pt}{\textbf{Task Types}} \\ \cline{2-7}
    ~ &\textbf{LoC} &\textbf{CC} &\textbf{NPath} &\textbf{Main}	&\textbf{Total} &\textbf{Depth} &~ \\ \hline
    
    userparallel    & 303   & 41    & 168     & 1		& 6		& 3		& 4 \\ \hline
    autoparallel 	& 406   & 43    & 344     & 2		& 7		& 3		& 20 \\ \hline
  \end{tabular}
  \caption{QR code comparison}
  \label{tab:qr_code}
\end{table}

\begin{figure}[!htb]
  \centering
  \includegraphics[width=0.9\linewidth]{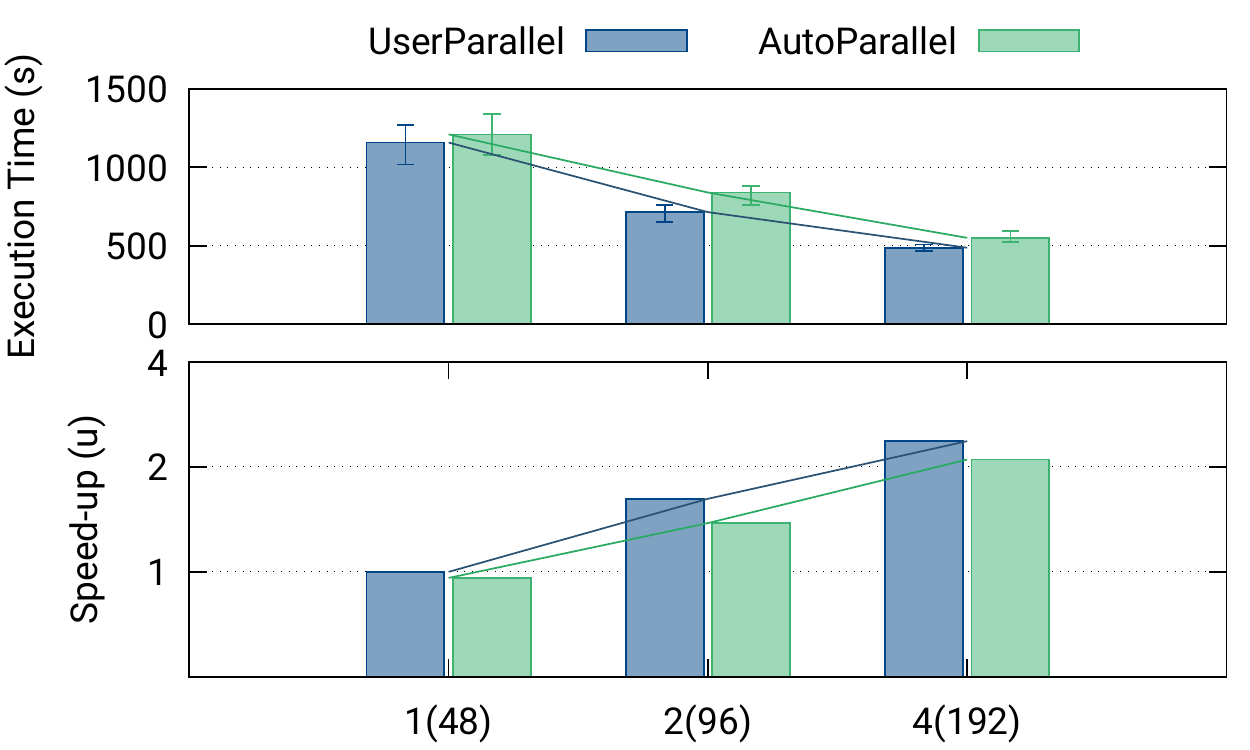}
  \caption{QR decomposition: Execution times and speed-up}
  \label{fig:qr_scaling}
\end{figure}

Figure~\ref{fig:qr_scaling} shows the execution results of the QR decomposition with a $32768 \times 32768$ matrix of $16 \times 16$ blocks with $2048 \times 2048$ elements each. The \textit{autoparallel} version spawns 26304 tasks and the \textit{userparallel} version spawns 19984 tasks. As in the previous examples, the overall performance is limited by the block size. However, the \textit{userparallel} version slightly outperforms the \textit{autoparallel} version; achieving a 2.37 speed-up with 4 workers instead of 2.10. The difference is mainly because the \textit{autoparallel} version spawns four copy tasks per iteration (\verb|copy_reference|), while the \textit{userparallel} version executes this code in the master side copying only the reference of a future object.

%--------------------------------
% Fine-grain Applications
\subsection{Fine-grain Applications}

The second set of experiments is designed to evaluate the capability of generating distributed code from a purely sequential code. To this end, we have implemented a Python version of many applications from the Polyhedral Benchmark suite~\cite{web:polybench}. 

As discussed in next subsections, this approach provides several advantages in terms of code re-organization and data blocking but, in its current state, has still significant performance issues. Hence, this paper only presents a preliminary evaluation of the GEMM application.

% GEMM
\subsubsection{GEMM} ~\newline

The presented implementation of the General Matrix-Matrix product of general rectangular matrices with float complex elements performs: $C = \alpha \cdot A \cdot B + \beta \cdot C$. In general terms, the arrays and matrices are implemented as plain NumPy arrays or matrices. This means that there are no \textit{blocks} and thus, the minimum work entity is a single element (a float). As in the previous set of experiments, the initialization is performed in a distributed way (defining tasks to initialize the matrix elements), and the computations are performed considering that the data is already located in a given node. 

We have evaluated a \textit{userparallel} Fine-Grain version and an \textit{autoparallel} version built using loop taskification. For comparison purposes, we have also evaluated a \textit{userparallel} Blocked version. Although the \textit{userparallel FG} works with single elements and \textit{userparallel B} with blocks, both versions include 2 tasks: \verb|scale|, and \verb|multiply|. The \textit{autoparallel LT} version defines four tasks: the two original ones and their two loop-tasked versions. The original tasks are kept because, in configurations that do not use PLUTO's tiles, it is possible to find function calls that cannot be loop-taskified. However, in this case, only the loop-tasked versions are called during the execution. Concerning the code, as shown in Table~\ref{tab:gemm_code}, the Loop Tasking version is significantly more complex in terms of lines of code, cyclomatic complexity, and n-path, but is capable of splitting the main loop into two loops (one for the scaling operations and another for the multiplications) for better parallelism.

\begin{table}[!htb]
  \centering
  \begin{tabular}{| m{1.4cm} | m{0.5cm} | m{0.5cm} | m{0.65cm} | m{0.5cm} | m{0.5cm} | m{0.6cm} | m{0.6cm} |}
    \hline
    \multirow{2}{*}{\textbf{Version}} & \multicolumn{3}{c|}{\textbf{Code Analysis}} & \multicolumn{3}{c|}{\textbf{Loops Analysis}} & \multirow{2}{0pt}{\textbf{Task Types}} \\ \cline{2-7}
    ~ &\textbf{LoC} &\textbf{CC} &\textbf{NPath} &\textbf{Main}	&\textbf{Total} &\textbf{Depth} &~ \\ \hline
    
    UserP. FG   & 194   & 22    & 112     & 1		& 4		& 3		& 2 \\ \hline
    UserP. B    & 189   & 22    & 112     & 2		& 5		& 3		& 2 \\ \hline
    AutoP. LT   & 382   & 133   & 360064    & 2		& 4		& 3		& 4 \\ \hline
  \end{tabular}
  \caption{Gemm code comparison}
  \label{tab:gemm_code}
\end{table}

\begin{figure}[!htb]
  \centering
  \includegraphics[width=0.7\linewidth]{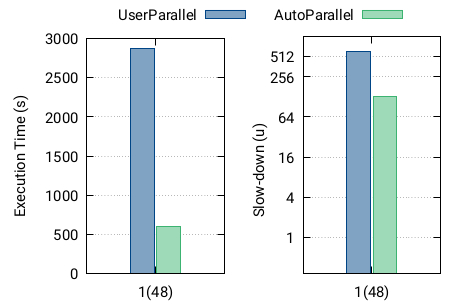}
  \caption{GEMM: Execution times and slow-down}
  \label{fig:gemm_scaling}
\end{figure}

Figure~\ref{fig:gemm_scaling} shows the execution results of the GEMM application with a matrix of $64 \times 64$ elements with 1 worker (48 cores). For the \textit{autoparallel}, the tile sizes are set to 8 and, for the blocked \textit{userparallel}, the matrix has $8 \times 8$ blocks with $8 \time 8$ elements each. The left plot shows the execution time of the \textit{userparallel FG} (blue) and the \textit{autoparallel LT} (green). The right plot shows the \textbf{slow-down} of both versions with respect to the blocked \textit{userparallel B} version running with a single worker.

Although both performances are completely unacceptable, there is still room for comparison. First, the \textit{autoparallel} version is capable of splitting the main loop to isolate the \verb|scaling| operations. Second, defining single elements as the minimum task entity leads to tasks with too little computation that cause a massive overhead of serialization and transfer inside PyCOMPSs. On the other hand, automatically building data blocks from the sequential user code improves almost 5 times the application performance. More in-depth, building data blocks helps PyCOMPSs to surpass the serialization and transfer overheads. However, building data blocks also increases significantly the number of parameters of each task (around $8 \times 8 \times 3 = 192 $ parameters per task) which slows down the PyCOMPSs scheduler and task manager.

We believe that Loop Tasking is a good approach for this kind of applications provided that PyCOMPSs is enhanced with support for collections. This means that task parameters should be defined as lists of objects that are handled by the PyCOMPSs runtime as a single entity. This would significantly lower the scheduling, serialization, and transferring overheads.

\section{Conclusions and Future Work}
\label{conclusions}

This paper has presented and evaluated AutoParallel, a Python module to automatically parallelize affine loop nests and execute them on distributed infrastructures. Built on top of PyCOMPSs and PLUTO, it is based on sequential programming so that anyone can scale up an application to hundreds of cores. Instead of manually taskifying a sequential python code, the users only need to add a \verb|@parallel| annotation to the methods containing affine loop nests.  

The evaluation shows that the codes automatically generated by the AutoParallel module for the Cholesky, LU, and QR applications can achieve the same performance than the manually parallelized versions. Thus, AutoParallel goes one step further in easing the development of distributed applications. 

As future work, although the Loop Taskification provides an automatic way to create blocks from sequential applications, its performance is still far from acceptable. On the one hand, we will do research on how to simplify the chunk accesses from the AutoParallel module. On the other hand, we will extend PyCOMPSs to support collection objects (e.g., lists) with different sizes so that we could avoid flattening and rebuilding of chunks, and serializing each object in a separate file.

Finally, AutoParallel could be integrated with different tools similar to PLUTO to support a larger scope of loop nests. For instance, APOLLO~\cite{apollo1, apollo2} provides automatic, dynamic and speculative parallelization and optimization of programs' loop nests of any kind (for, while or do-while loops). However, its integration would require PyCOMPSs to be extended with some speculative mechanisms.

% ACKNOWLEDGMENTS
\ifCLASSOPTIONcompsoc
  \section*{Acknowledgments}
\else
  \section*{Acknowledgment}
\fi
This work has been supported by the Spanish Government (SEV2015-0493), by the Spanish
Ministry of Science and Innovation (contract TIN2015-65316-P), by Generalitat de Catalunya (contract 2014-SGR-1051), and by the European Commission through the Horizon 2020 Research and Innovation program under contract 800898 (ExaQUte project). Cristian Ramon-Cortes predoctoral contract is financed by the Ministry of Economy and Competitiveness under the contract BES-2016-076791. 
% This work has partially been supported by the Joint Laboratory for Extreme Scale Computing (JLESC)
%This work has been supported by the European Commission through the Horizon 2020 Research and Innovation program under contract 800898 (ExaQUte project).

Thanks to C\'edric Bastoul for his support with CLooG.


% REFERENCES SECTION

%\bibliographystyle{IEEEtran}
\bibliographystyle{IEEEtran}
\bibliography{00autoparallel_main}

% APPENDIX: Moved to a separated file
\clearpage
\begin{appendices}
    \section{Artifact Description}

%----------------------------------------------------------------------------------------
\subsection{Abstract}
This description contains the information required to launch the experiments of the SC18 paper "AutoParallel: A Python module for automatic parallelization and distributed  execution  of  affine  loop  nests". More precisely, we explain how to install the AutoParallel module and its dependencies, and how to run the experiments described in Section~\ref{experimentation}. 

%----------------------------------------------------------------------------------------
\subsection{Description}

\subsubsection{Check-list (artifact meta information)}
\begin{itemize}
    \item Program: Python application, Python binding, Java Runtime, C and C++ libraries
    \item Run-time environment: AutoParallel 0.2, PyCOMPSs 2.3.rc1807, PLUTO 0.11.4, Python 2.7.13, Java OpenJDK 8 131, GCC 7.2.0, Boost 1.64.0
    \item Output: Time to solution and number of tasks
    \item Experiment workflow: Prepare system, clone PyCOMPSs, load submodules, install PyCOMPSs, install PLUTO, install the AutoParallel module, run the examples, and observe the results
    \item Experiment customization: Input dataset size, number of nodes, number of cores, maximum node memory, PLUTO tile sizes, and PyCOMPSs scheduler, log level, communication adaptor, and workers working directory
    \item Publicly available?: Yes
\end{itemize}

\subsubsection{How delivered}
PyCOMPSs and the AutoParallel module can be cloned from GitHub using~\cite{github:compss} and~\cite{github:autoparallel} respectively. The examples used for the experimentation can be found under the \verb|examples| folder. 

\subsubsection{Hardware dependencies}
None. 

\subsubsection{Software dependencies}
PyCOMPSs depends on the COMPSs Runtime, the Python and Commons bindings, and the Extrae tracing module that are automatically installed. However, the users must provide valid Java, Python, and GCC installations.

PLUTO depends on Candl, Clan, CLooG, ISL, OpenScop, PipLib, and PolyLib modules that are automatically installed. However, the users must explicitly provide valid GMP, Flex, and Bison installations. 

The AutoParallel module requires a valid installation of PyCOMPSs and PLUTO. Moreover, the users must check that the following Python modules are available: AST, ASTOR, enum34, logging, inspect, islPy, symPy, subprocess, and unittest.

The examples used for the experimentation require the NumPy Python module. 

\subsubsection{Datasets}
The datasets of each application are pseudo-randomly generated during the execution since the size is the only relevant parameter for the computational results.
However, the users can check the correctness of the algorithms against the their sequential version by using the \textit{-d} option when invoking the run scripts.

%----------------------------------------------------------------------------------------
\subsection{Installation}
\begin{enumerate}
    \item Prepare the system
        \begin{lstlisting}[style=Bash, basicstyle=\scriptsize]
# Runtime dependencies
$ sudo apt-get install openjdk-8-jdk uuid-runtime curl wget openssh-server maven graphviz xdg-utils
$ export JAVA_HOME=<path_to_openjdk>
# Python and Commons bindings dependencies
$ sudo apt-get install libtool automake build-essential python-dev libpython2.7 python-pip libboost-all-dev libxml2-dev csh
$ sudo pip2 install numpy dill decorator
# Extrae tracing module dependencies
$ sudo apt-get install libxml2 gfortran libpapi-dev papi-tools
# AutoParallel and PLUTO dependencies
$ sudo apt-get install libgmp3-dev flex bison libbison-dev texinfo
$ sudo pip2 install astor enum34 islpy sympy
        \end{lstlisting}
        
    \item Clone PyCOMPSs and enter the newly created directory
        \begin{lstlisting}[style=Bash, basicstyle=\scriptsize]
$ git clone https://github.com/bsc-wdc/compss.git
$ cd compss
        \end{lstlisting}
        
    \item Initialize and patch the submodules (PLUTO, AutoParallel, and Extrae)
        \begin{lstlisting}[style=Bash, basicstyle=\scriptsize]
$ ./submodules_get.sh
$ ./submodules_patch.sh
        \end{lstlisting}
        
    \item Install everything (a) into a given target location or (b) into the default location \verb|/opt/COMPSs| (requires root privileges)
            \begin{lstlisting}[style=Bash, basicstyle=\scriptsize]
$ cd builders
$ (a) ./buildlocal $HOME/COMPSs     
$ (b) sudo -E ./buildlocal
\end{lstlisting}
    \item Check up the environment
        \begin{lstlisting}[style=Bash, basicstyle=\scriptsize]
$ runcompss -v
COMPSs version 2.3 Daisy
        \end{lstlisting}
\end{enumerate}

%----------------------------------------------------------------------------------------
\subsection{Experiment workflow}
Once that PyCOMPSs, PLUTO, and the AutoParallel module are installed, any example can be executed. To ensure that everything runs smoothly, the users should use the prepared scripts: the \verb|run.sh| for laptops and the \verb|enqueue.sh|, and \verb|experiments.sh| scripts for supercomputers. 

More in detail, each application contains:
\begin{itemize}
    \item \verb|README.md| : File describing the application and its commands 
    \item \verb|autoparallel| : Folder containing the autoparallel version of the application
        \begin{itemize}
            \item \verb|app_name.py| : Source file of the autoparallel version of the application
            \item \verb|app_name_autogen.py| : Source file automatically generated 
            \item \verb|run.sh| : Script to run the autoparallel version of the application
        \end{itemize}
    \item \verb|userparallel| : Folder containing the userparallel version of the application
        \begin{itemize}
            \item \verb|app_name.py| : Source file of the userparallel version of the application
            \item \verb|run.sh| : Script to run the userparallel version of the application
        \end{itemize}
    \item \verb|run.sh| : Script to run the all the versions of the application
    \item \verb|enqueue.sh| : Script to enqueue the application to a queue system (SLURM, LSF, or PBS)
    \item \verb|experiments.sh| : Script to run all the experiments with all the versions of the application
    \item \verb|results.sh| : Script to parse the experiments results
\end{itemize}

In the rest of the artifact description, we will explain the most important options to set up in order to reproduce the results of the paper.

%----------------------------------------------------------------------------------------
\subsection{Evaluation and expected result}

% LOCAL
On the one hand, the \verb|run.sh| script is prepared to run executions on laptops. Although it can be simply invoked without parameters (setting up the default values), the script is a wrapper of the \verb|runcompss| command and accepts many additional flags (e.g., \verb|-d| for debug). For instance, here is the command to execute the Cholesky application:
\begin{lstlisting}[style=Bash, basicstyle=\scriptsize]
$ cd examples/cholesky
$ ./run.sh
\end{lstlisting}

Two executions will be triggered; one for the \textit{autoparallel} and another for the \textit{userparallel} version. The output for the Cholesky example should include the following lines:

\begin{lstlisting}[style=Bash, basicstyle=\scriptsize]
RESULTS -----------------
VERSION AUTOPARALLEL
MSIZE 4
BSIZE 4
DEBUG False
TOTAL_TIME 3.91822195053
INIT_TIME 3.22278118134
CHOLESKY_TIME 0.695440769196
-------------------------
\end{lstlisting}
\begin{lstlisting}[style=Bash, basicstyle=\scriptsize]
[(5330) API] - -- COMPSs Task Execution Summary --
...
[(5331) API] - Total executed tasks: 36
[(5331) API] - -----------------------------------
\end{lstlisting}

The results section provides information about the application execution, such as the application parameters and the computation time (\verb|CHOKESKY_TIME| in the example). The execution summary shows, between others, the total number of executed tasks. Furthermore, inside the results folder the users will find the paraver traces and the task dependency graphs:

\begin{lstlisting}[style=Bash, basicstyle=\scriptsize]
$ cd examples/cholesky/results/local/
$ okular autoparallel/complete_graph.pdf
$ okular userparallel/complete_graph.pdf 
$ wxparaver autoparallel/trace/*.prv
$ wxparaver userparallel/trace/*.prv
\end{lstlisting}

% QUEUES
On the other hand, the \verb|enqueue.sh|, and \verb|experiments.sh| scripts are prepared to run executions on supercomputers. The first is a wrapper of the \verb|enqueue_compss| to enqueue a single execution of the application. The users must specify version (autoparallel or userparallel), job dependency, number of nodes, wall-clock time, number of cpus per node, tracing value, graph value, log level, and the application parameters (in the Cholesky example: the matrix size and block size). 
\begin{lstlisting}[style=Bash, basicstyle=\scriptsize]
$ ./enqueue.sh autoparallel None 2 150 48 true false off 32 2048
\end{lstlisting}

The \verb|experiments.sh| script enqueues several jobs to reproduce the experimentation described in Section~\ref{experimentation} and can be run without arguments. The results can be easily summarized to a CSV file by running the \verb|results.sh| script, that will create a \verb|results| folder with a \verb|results.summary| file inside containing:
\begin{lstlisting}[style=Bash, basicstyle=\scriptsize]
JOB_ID	VERSION	MSIZE	BSIZE	TRACING	NUM_WORKERS	TOTAL_TIME	INIT_TIME	COMP_TIME	NUM_TASKS
2210018	autoparallel	32	2048	true	1	520,067487001	93,8065400124	426,260946989	7008
\end{lstlisting}

%----------------------------------------------------------------------------------------
\subsection{Experiments Customization}

The application parameters can be modified inside the corresponding \verb|experiments.sh| script of each application. Moreover, the PyCOMPSs options can be tuned inside the corresponding \verb|enqueue.sh| script. 

%----------------------------------------------------------------------------------------
%\subsection{Final Notes}
% NONE
\end{appendices}

\end{document}